\documentclass[twocolumn,showpacs,prl]{revtex4}

\usepackage[dvips]{graphicx,color}
\usepackage{delarray}
\usepackage{amsmath}
\usepackage{bm}

\def \binom#1#2{{#1\choose #2}}

\begin{document}
\title{Measurement-device-independent quantum key distribution with all-photonic adaptive Bell measurement}

\author{Koji Azuma}
\email{azuma.koji@lab.ntt.co.jp}
\affiliation{NTT Basic Research Laboratories, NTT Corporation, 3-1 Morinosato Wakamiya, Atsugi, Kanagawa 243-0198, Japan}

\author{Kiyoshi Tamaki}
\affiliation{NTT Basic Research Laboratories, NTT Corporation, 3-1 Morinosato Wakamiya, Atsugi, Kanagawa 243-0198, Japan}

\author{William J. Munro}
\affiliation{NTT Basic Research Laboratories, NTT Corporation, 3-1 Morinosato Wakamiya, Atsugi, Kanagawa 243-0198, Japan}

\
\date{\today}

\begin{abstract}
The time-reversed version of entanglement-based quantum key distribution (QKD), called measurement-device-independent QKD (mdiQKD), was originally introduced to close arbitrary security loopholes of measurement devices. Here we show that the mdiQKD has another advantage which should be distinguished from the entanglement-based QKD. In particular, an all-photonic adaptive Bell measurement, based on the concept of quantum repeaters, can be installed solely in the mdiQKD, which  leads to a square root improvement in the key rate. 
This Bell measurement also provides a similar improvement in the single-photon-based entanglement generation of quantum repeaters.
\pacs{03.67.Dd, 03.67.Hk, 03.65.Ud, 42.50.Ex}
\end{abstract}
\maketitle

Quantum key distribution (QKD) enables two distant legitimate parties, called Alice and Bob, 
to share a secret key under the intervention of an eavesdropper, called Eve \cite{review1,review2}.
QKD is definitely a leading technology in quantum information science. In fact, QKD networks have been demonstrated over fiber networks in the environment and even commercial products exist \cite{review1,review2}. However, unfortunately, the implementations have a fundamental limitation on the achievable communication rates, owing to the photon loss that increases exponentially with the communication distance. In principle, quantum repeaters \cite{B98,DLCZ,SSRG09,KWD03,C06,S09,ATKI10,ZDB12,M10,M12,L12,G12,AK12,J09,L08,R09,ATL13,L06,M08,A09,ASKI10} solve this problem and allow us to perform QKD over arbitrary long distances efficiently. However, there is actually a large gap between the present technologies and ones assumed in quantum repeaters. 
Hence, a protocol to bridge this gap is essential for developing QKD networks seamlessly, which we will present in this paper.

The QKD protocol was first proposed by Bennett and Brassard \cite{BB84}, referred to as BB84.
In this protocol, Bob performs measurements in complementary bases, $Z$-basis and $X$-basis, on photons from Alice.
Bennett {\it et al.} relate this prepare-and-measure protocol with an entanglement-based QKD proposed by Ekert \cite{E91}, 
by regarding the BB84 as a modified entanglement-based one where Alice and Bob perform measurements in the complementary bases on Bell pairs \cite{BBM92}.
Based on this equivalence of these protocols, they are similarly shown to be unconditionally secure in principle \cite{review1,review2}.
However, in the practical implementations, there are possibilities that the physical devices of Alice and Bob do not work as the security proofs require. In particular, measurement devices used in those implementations may have security loopholes that can be opened by optical pulses from Eve and are exploitable for her attacks.
In fact, most successful attacks on QKD utilized this kind of loopholes of practical photon detectors \cite{detectoratt,Z08,L10,G11}.

To close all the loopholes of measurement devices, Lo {\it et al.} proposed the concept of measurement-device-independent QKD (mdiQKD) \cite{LCQ12}, which is the time-reversed version of the entanglement-based QKD protocol.
In fact, in the entanglement-based QKD, Alice and Bob respectively perform measurements on photon pairs that may have been prepared in a Bell state and distributed by a node between them, while, in the mdiQKD, Alice and Bob independently prepare photons and send them to the node that is supposed to apply Bell measurements to the received pairs.
This time reversal makes a difference in the assumption on which physical devices are reliable. 
In particular, the security of the former relies on the measurement devices of Alice and Bob alone, 
while that of the latter does only on their sending devices. 
As a result, the mdiQKD can be secure without putting any assumption on the measurement devices.
However, except for this difference, the other aspects such as the security proofs and the key rates are essentially the same, owing to the simple time-reversed nature.

In this paper, we show, contrary to what one may infer from those similarities, that the mdiQKD has another significant advantage which cannot be seen in entanglement-based QKD.
In particular, we utilize a distinguished feature of the mdiQKD that the intermediate node receiving photons from Alice and Bob has an option to perform Bell measurements only on surviving photons under losses. This {\it adaptive Bell measurement} leads to a {\it square root improvement} in the key rate, thanks to the concept of quantum repeaters behind it. 
The effectiveness of combining this repeater concept with the mdiQKD has first been suggested by Panayi {\it et al.} \cite{PRML14}.
However, their protocol uses matter quantum memories of conventional quantum repeaters
\cite{DLCZ,SSRG09}. In contrast, our adaptive Bell measurement follows an ``all photonic approach'' presented in Ref.~\cite{ATL13} for quantum repeaters. As a result, it can be realized only with optical switches, single-photon sources, photon detectors, and active feed-forward techniques, reaping the following benefits resulting from removing the necessity of matter quantum memories \cite{ATL13}: (a) No memory implies that the repetition rate can be increased as high as one wants within those allowed by assumed devices, shared with memory-function-less quantum repeaters \cite{M12}. (b) Even if we realize a single-photon source with a matter qubit, the matter
qubit is not needed to have a deterministic interaction with photons as well as to have long
coherence time (and, of course, a matter quantum memory \cite{SSRG09,S10r} can be diverted to a single-photon source).
(c) Coherent frequency converters for photons to strengthen the coupling to matter quantum memories \cite{T05} and to optical fibers \cite{I11} could be unnecessary.
(d) Our protocol could work at room temperature in principle.
In addition, the adaptive Bell measurement also provides a similar improvement in the single-photon-based entanglement generation process of quantum repeaters \cite{DLCZ,SSRG09}.
Therefore, our scheme will play an important role to bridge the gap between QKD and quantum repeaters, technologically and conceptually.

\begin{figure}[b]  
\includegraphics[keepaspectratio=true,height=33mm]{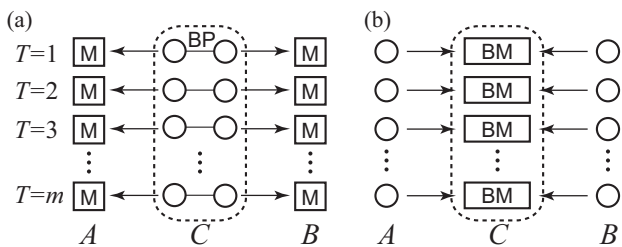}
  \caption{Entanglement-based QKD and mdiQKD. $T$ is the number of trials. 
(a) In the entanglement-based QKD, the node $C$ sends halves of Bell pairs (BP) to Alice and Bob who just perform measurements (M), respectively. (b) In the mdiQKD, the node $C$ performs Bell measurements (BM) on photons that have been sent by Alice and Bob. 
}
\label{fig:1.eps}
\end{figure}
 
{\it Entanglement-based QKD and mdiQKD.}---It is instructive to start by reviewing the relation between the entanglement-based QKD and the mdiQKD in more detail (Fig.~\ref{fig:1.eps}). 
Suppose that Alice and Bob are separated over distance $L$, and a node $C$ between them shares optical channels with them.
For simplicity, the node $C$ is assumed to be located in the middle of Alice and Bob.
Then, the transmittance of the optical channels is described by $e^{-L/(2l_{\rm att})}=:\eta_{L/2}$ with an attenuation length $l_{\rm att}$.
The transmittance is the same as the arrival probability of a single photon through the lossy channels.
In the entanglement-based protocol [Fig.~\ref{fig:1.eps}(a)], the node $C$ respectively sends the halves of a photonic Bell pair to Alice and Bob via the optical channels, each of which is subject to a $Z$-basis or $X$-basis measurement. Since Alice and Bob keep the measurement outcomes for the shifted key only when both of them find the arrivals of photons with their measurements,
the probability $P^{\rm sif}$ of keeping the outcomes scales with $\eta_{L/2}^2=\eta_L$ in practice.
On the other hand, in the mdiQKD [Fig.~\ref{fig:1.eps}(b)], instead of sending Bell pairs as in the entanglement-based QKD, the node performs a Bell measurement to a pair of single photons that have been prepared randomly in one of the eigenstates of complementary observables $\hat{Z}$ and $\hat{X}$ and sent simultaneously by Alice and Bob.
This is the time reversal of the entanglement-based protocol, and it thus follows the same scaling, i.e., $P^{\rm sif} \sim \eta_L$.

The probability $P^{\rm sif} \sim \eta_{L}$ for the entanglement-based QKD and the mdiQKD implies that the number of trials required to obtain a pair of bits for the shifted key is $\eta_L^{-1}$ on average.
This scaling is shared with all conventional protocols including prepare-and-measure QKD \cite{review1,review2}.
To improve the scaling $\eta_L^{-1}$ to $\eta_{L}^{-1/2}(=\eta_{L/2}^{-1})$, Panayi {\it el al.} introduce matter quantum memories to the node $C$ in the mdiQKD setting \cite{PRML14}. 
However, as we will show, the matter quantum memories are not necessary for this improvement.

It is essential for our {\it all-photonic} approach to notice that the original scaling $\eta_L^{-1}$ is caused by a fact that the pairings at the node $C$ for Bell pairs in the entanglement-based QKD and for Bell measurements in the mdiQKD are {\it predetermined independently of the existence of photon losses}. 
In other words, to outperform the $\eta_L^{-1}$ scaling, we need to make the pairings depend on the occurrences of photon losses.
Interestingly, this is possible {\it solely} for the mdiQKD, because it entangles photons {\it after} the transmissions in contrast to the entanglement-based QKD (c.f., Fig.~\ref{fig:1.eps}). 

\begin{figure}[b]  
\includegraphics[keepaspectratio=true,height=32mm]{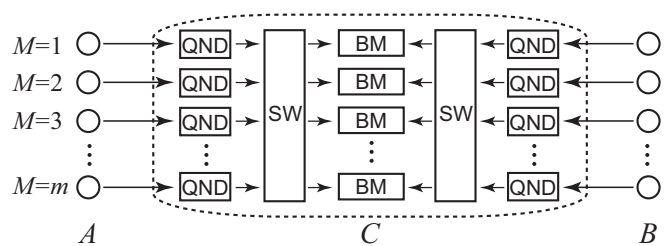}
  \caption{Basic idea of our mdiQKD protocol with an adaptive Bell measurement. $M$ is the number of pulses. 
In this protocol, the node $C$ first performs quantum non-demolition (QND) measurements to confirm the successful arrival of single photons, followed by optical switches (SW) to send the surviving photons to Bell measurement (BM) modules.}
\label{fig:2.eps}
\end{figure}

{\it Basic idea of adaptive mdiQKD.}---To make our statement more precise,
we introduce our mdiQKD protocol (Fig.~\ref{fig:2.eps})  where the node $C$ performs an adaptive Bell measurement.
This protocol proceeds as follows:
(i) Alice and Bob send $m$ pulses in single-photon states, randomly selected from the eigenstates of complementary observables, to the node $C$ {\it at the same time},
using multiplexing.
(ii) On receiving the pulses, the node $C$ applies quantum non-demolition (QND) measurements to the pulses in order to confirm the arrival of the single photons over lossy channels.
(iii) Then, successfully arriving photons from Alice are paired with ones from Bob via optical switches at the node $C$.
(iv) The node $C$ then performs a Bell measurement on each of these pairs.
(v) The node $C$ then announces the pairings and the measurement outcomes of the Bell measurements.
(vi) Finally, as bits for the shifted key, Alice and Bob keep the eigenvalues corresponding to their sent eigenstates to which the Bell measurements have been successfully applied.

Let us consider the scaling of our protocol with distance $L$. 
When Alice and Bob's pulses are perfectly in single-photon states, the transmittance $\eta_{L/2}$ of the channels affects only to the probability of 
confirming the arrival of single photons via QND measurements in step (ii). 
Since this probability is proportional to $\eta_{L/2}$, 
if the number $m$ of multiplexing is larger than $\eta_{L/2}^{-1}$, some single photons (almost) deterministically arrive at the node $C$ from both Alice and Bob on average.
Since the successful application of the Bell measurement to these single photons leads to a pair of bits for the shifted key in step (vi),
the resources to obtain the one pair are in the order of $\eta_{L/2}^{-1}$.
This is a square root improvement over the original mdiQKD and comes from making the pairings for the Bell measurement depend on the successful arrival of single photons at node $C$.

The precise performance of our protocol can be evaluated as follows.
Let us assume that Alice and Bob have photon sources with efficiency $\eta_{\rm s}$.
Now suppose that the QND measurements in step (ii) and the Bell measurements in step (iv) succeed with probabilities $p_{\rm QND}$ and $p_{\rm BM}$, respectively.
The probability $p_{k|m}$ with which node $C$ finds the existence of $k(\le m)$ single photons from Alice or from Bob via QND measurements in step (ii) is
\begin{equation}
p_{k|m}=B_{k|m}(p_{\rm QND} \eta_{L/2} \eta_{\rm s} ),
\end{equation}
where $B_{k|m}(p)$ is the binomial distribution with 
$
B_{k|m}(p):=
\binom mk p^k (1-p)^{m-k}.
$
To make $l$ pairs in step (iii), the node $C$ should have found the existence of single photons not less than $l$ from both of Alice and Bob in step (ii).
Hence, the probability $P^{\rm sif}_{n|m}$ with which our protocol provides $n$ pairs of bits for the shifted key in step (vi) is described as
\begin{equation}
P^{\rm sif}_{n|m}=\sum_{l=n}^m B_{n|l}(p_{\rm BM}) f_{l|m}
\end{equation}
with probability distribution
$
 f_{l|m}:=  2 p_{l|m} \sum_{k=l}^m p_{k|m}
-p^2_{l|m} 
$.
The average number $\bar{n}_m$ of shifted pairs is then
\begin{align}
\bar{n}_m =& \sum_{n=0}^m n P^{\rm sif}_{n|m}=p_{\rm BM} \sum_{l=0}^m     f_{l|m} l \nonumber \\
=&m p_{\rm BM}[  p_{\rm QND} \eta_{L/2} \eta_{\rm s}-  g_m(p_{\rm QND} \eta_{L/2} \eta_{\rm s})],
\label{eq:n}
\end{align}
where $g_m$ is shown \cite{g_m} to be
\begin{multline}
g_m(p)=p(1-p)\\
\times \Biggl[ \sum_{l=0}^{m-1} B_{l|m-1}^2(p) 
+ \sum_{l=1}^{m-1} B_{l|m-1}(p) B_{l-1|m-1}(p)\Biggr]. \label{eq:g_m}
\end{multline}
Since the maximum of $B_{l|m-1}(p)$ over $l$ goes to zero in the limit of $m\to \infty$, 
we have $\lim_{m\to \infty} g_{m}=0$. Therefore, the asymptotic shifted-key generation rate $R:=\lim_{m\to \infty} \bar{n}_m/m $ is 
\begin{equation}
R=  p_{\rm BM}  p_{\rm QND} \eta_{L/2} \eta_{\rm s}. \label{eq:R}
\end{equation}
Since the rate of the original mdiQKD protocol is described by $R= p_{\rm BM} \eta_{L}  \eta_{\rm s} $, our protocol necessitates, at least, 
\begin{equation}
p_{\rm QND}>\eta_{L/2} \label{eq:cond}
\end{equation}
to outperform the original mdiQKD protocol in terms of $R$.
Equation~(\ref{eq:R}) also shows that the multiplexing number $m$ should be in the order of $m\sim (p_{\rm BM}  p_{\rm QND} \eta_{L/2} \eta_{\rm s} )^{-1} $ to obtain one pair of bits.

{\it All photonic implementation.}---We present an example of all photonic realization of our mdiQKD. The Bell measurement in step (iv) can be executed just by using linear optical elements and photon detectors \cite{K07}.
The most challenging technique in our mdiQKD is the QND measurement in step (ii).
Fortunately, there are several all-photonic realizations of QND measurements for single photons \cite{K07}.
Here we focus on a simple example, i.e., a QND measurement for a single photon \cite{K02} based on quantum teleportation \cite{B93}.
This scheme teleports the single-photon state of the incoming pulse to that of a half of a photonic Bell pair via the linear-optics-based Bell measurement,
utilizing its feature that the teleportation fails when the incoming pulse is in the vacuum state. Thus, if the received pulse is in a single-photon state, the success of this teleportation confirms the existence of a single photon in the pulse without disturbing the single-photon state.

Assuming the all-photonic QND measurement based on quantum teleportation for an implementation of our mdiQKD, 
we can now estimate the final key rate $G$ per pulse (normalized by the number of events of the same basis choice by Alice and Bob) with the asymptotic formula \cite{review1}
\begin{equation}
G= R [1-h(e_Z)-h(e_X)],
\end{equation}
where $e_Z$ is the bit error rate, $e_X$ is the phase error rate, and $h(x)=-x\log_2 x-(1-x)\log_2 (1-x)$.
$R$ depends only on the success probabilities of assumed processes in a protocol as described by Eq.~(\ref{eq:R}). 
If we regard Alice and Bob's single photons in step (i) as ones entangled with their own virtual qubits, mdiQKD protocols can be considered to be the process for entangling their virtual qubits \cite{BBM92,LCQ12}. In this paradigm, $e_Z$ and $e_X$ represent the quality of the entanglement of the virtual qubits.
Hence, if we specify devices for our mdiQKD, we can estimate the final key $G$.

For simplicity, we assume single-photon sources with efficiency $\eta_{\rm s}$ and with pulse width $\tau_{\rm s}$, and single-photon detectors with quantum efficiency $\eta_{\rm d}$ and with mean dark count rate $\nu_{\rm d}$.
Let $\eta_{\rm s}=0.90$ \cite{SSRG09}, $\tau_{\rm s}=300$~ps, $\eta_{\rm d}=0.93$, and $\nu_{\rm d}=1$~s$^{-1}$ \cite{PRML14,M13,S10}.
Our protocol needs an active feedforward technique with an optical switch.
Suppose that a single active feedforward can be completed within time $\tau_{\rm a}$, during which photons run in optical fibers.
Let $\tau_{a}=150$~ns \cite{P07} and $l_{\rm att}=22$~km for all optical fibers, and assume that the speed of light in optical fibers is $c=2.0 \times 10^8$~m/s. 
Bell pairs for the all-photonic QND measurements in step (ii) can be generated in constant time $\tau_{\rm a}$ with single-photon sources and the active feedforward with switching. 
In fact, a Bell pair can be generated with linear optical elements and single-photon sources \cite{BR05} only probabilistically,  but, if we parallelize this probabilistic generation process so that we can obtain, at least, a Bell pair, we are (almost) always able to pick up the Bell pair via a single application of the active feedforward with switching.
In practice, this kind of step-wise preparation of Bell pairs can suppress the multi-photon emissions, which would be necessary for satisfying Eq.~(\ref{eq:cond}) with similar reasoning of Ref.~\cite{PRP14} concerning the scheme \cite{PRML14} of Panayi {\it et al}.
In addition, we need to use one active feedforward in step (iii).

\begin{figure}[t]  
\includegraphics[keepaspectratio=true,height=44mm]{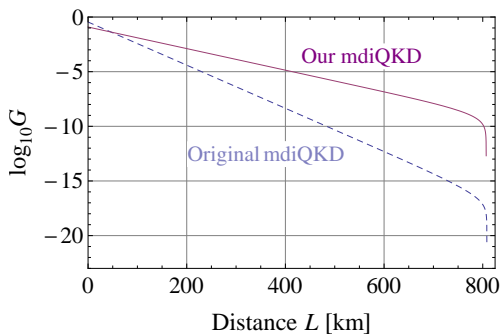}
  \caption{Secret key rates $G$ per pulse (normalized by the number of events of the same basis choice by Alice and Bob) versus distances $L$. Key rates $G$ for the original mdiQKD protocol with the same single-photon sources are also shown by the dashed curve as a reference. }
  \label{fig:3.eps}
\end{figure}

Under these assumptions, the final key rates $G$ are given as in Fig.~\ref{fig:3.eps}.
The figure shows that our protocol outperforms the original mdiQKD protocol in the region of long distances.
If single-photon sources with repetition rate of 1~GHz are available as assumed in Ref.~\cite{SSRG09}, the key generated per second is 0.15~Hz even for $L=800$~km, which is one order of magnitude better than the best scheme \cite{S07} of quantum repeaters with atomic ensembles \cite{SSRG09}.

{\it Applications.}---The adaptive Bell measurement is also useful for increasing the performance for any protocol based on the single-photon-based entanglement generation.
For instance, it can be installed in quantum repeaters with atomic ensembles \cite{SSRG09}, because they are based on such an entanglement generation process.
However, note that it is impossible to accomplish {\it all photonic quantum repeaters} as in Ref.~\cite{ATL13} with the adaptive Bell measurement alone.
In fact, although we can use our mdiQKD protocol as the entanglement generation for Alice and Bob by regarding their virtual qubits as actual qubits, they need to wait the arrival of the heralding signals from node $C$ in step (v) in order to identify the qubits that have successfully been entangled, which is impossible without the memory function of their qubits. 
Therefore, for extremely long distances such as thousand kilometers, quantum repeaters are needed.

In conclusion, we have presented an all-photonic adaptive mdiQKD protocol that can present a square root improvement over conventional QKD protocols. 
In our analysis, we have assumed the use of single-photon sources and single-photon detectors for simplicity. 
However, as long as the success probability of the QND measurement can be made independent of the distance between the sender and the repeater node, 
since the scaling of $R$ with distance $L$ does not change, 
our protocol could work with more practical devices.
For moderate communication distances, our protocol could work more efficiently than quantum repeaters with atomic-ensemble quantum memories \cite{SSRG09}.
In addition, our adaptive Bell measurement can be installed in any protocol using single-photon-based entanglement generation in order to improve the performance.
Combined with all photonic quantum repeaters \cite{ATL13}, our result paves a seamless route toward long-distance quantum communications with optical devices alone.

We thank G.~Kato and F.~Morikoshi, and especially M.~Curty, H.-K.~Lo, and N.~L\"{u}tkenhaus for valuable discussions.
This research is in part supported by the Project UQCC by the
National Institute of Information and Communications Technology (NICT).


\begin{thebibliography}{99}
\bibitem{g_m}
$g_m$ is defined by 
$
g_m(p):= m^{-1} ( \sum_{l=1}^m \sum_{k=l+1}^{l-1}  l f_{l|m}  f_{k|m}
-\sum_{l=1}^{m-1} \sum_{k=l+1}^{m} l f_{l|m}  f_{k|m} ) $. Eq.~(\ref{eq:g_m}) is derived by using
$
l B_{l|m}(q)= mqB_{l-1|m-1}(q)
$
for $l>0$ and 
$
B_{k|m}(q)= (1-q)B_{k|m-1}(q)+qB_{k-1|m-1}(q) 
$
for $0<k<m$.
\end{thebibliography}
\end{document}